\documentclass{aa}
\usepackage{graphicx,epstopdf}
\usepackage{amsmath}
\usepackage{amssymb}
\usepackage[svgnames,hyperref]{xcolor}
\definecolor{yourcolor}{RGB}{25,84,149} %Here you have the option to change the color of the citations/urls etc. to whatever color you like (in RGB code). If you just want blue, remove line 8 and use line 9 instead
\usepackage[breaklinks=true, colorlinks, linkcolor=yourcolor, citecolor=yourcolor, urlcolor=yourcolor]{hyperref}
\usepackage{url}
\usepackage{txfonts}

\def\ergcmsqs{erg\,cm$^{-2}$\,s$^{-1}$}

\def\ppc{pc$^{-3}$}
\def\ergs{erg\,s$^{-1}$}

\begin{document}

\title{On the space density of X-ray selected cataclysmic variables}
\author{A.D. Schwope\inst{1}}
\institute{Leibniz-Institut f\"ur Astrophysik Potsdam (AIP), An der Sternwarte
  16, 14482 Potsdam, Germany}
\date{Received ; accepted } 

\abstract{The space density of the various classes of cataclysmic variables (CVs) 
could only be weakly constrained in the past. Reasons were the small number of objects in
  complete X-ray flux-limited samples and the difficulty to derive precise 
  distances to CVs. The former limitation still exists. Here the impact of Gaia parallaxes and implied distances on the space density of X-ray selected complete, flux-limited samples is studied. The samples are described in the literature, those of non-magnetic CVs are based on 
ROSAT (RBS -- ROSAT Bright Survey \& NEP -- North Ecliptic Pole), 
that of the Intermediate Polars stems from Swift/BAT. All CVs appear to be rarer 
than previously thought, although the new values are all within the errors of past studies. 
Upper limits at 90\%  confidence for the space densities of non-magnetic CVs are 
$\rho_{\rm RBS} < 1.1 \times 10^{-6}$\,\ppc\, and 
$\rho_{\rm RBS+NEP} < 5.1 \times 10^{-6}$\,\ppc\, for an assumed scale height of $h=260$\,pc
and $\rho_{\rm IPs} < 1.3 \times 10^{-7}$\,\ppc\, for the long-period Intermediate Polars at 
a scale height of 120\,pc. Most of the distances to the IPs were under-estimated in the past. 
The upper limits to the space densities are only valid in the case where CVs
do not have lower X-ray luminosities than the lowest-luminosity member of the
sample.
These  results need consolidation by larger sample sizes, soon to 
be established through sensitive X-ray all-sky surveys to be performed with eROSITA on 
the Spektrum-X-Gamma mission.
}


\maketitle

\section{Introduction}
The space density of cataclysmic variable stars (CVs) is one of the bigger
unknowns in the field. Being the outcome of binary star evolution through a
common envelope and subsequent angular momentum loss this number is
important for several parameters and processes that are relevant in
binary evolution: the initial binary separation, the initial mass
distribution, the common envelope efficiency, the angular momentum loss in the
post-common envelope phase and that in the CV stage. It is the most relevant
number to compare with binary population synthesis. The simple
question is: How many are out there? 

The question is relevant for stellar evolution but has
implications for models of the total energy output of the Milky Way. One of the
contenders to explain the infamous Galactic Ridge X-ray Emission~\citep[GRXE,][]{worrall+82} are CVs and in particular magnetic CVs of the Intermediate
Polar type (IPs). Through a deep Chandra pointing close to the galactic
centre the apparently diffuse GRXE was largely resolved into point sources
~\citep{revnivtsev+09}. The composition however remained uncertain and was
discussed thoroughly in the recent past and depends 
sensitively on the space density and the luminosity functions of the
main source classes~\citep[e.g.][]{warwick14,nobukawa+16}.

The second data release from the Gaia satellite opens the opportunity to 
re-assess the space density of X-ray selected cataclysmic variables. Past studies 
were hampered by small sample sizes and imprecisely determined distances.
While the former limitation cannot be overcome presently the latter has 
essentially vanished.

In this paper the non-magnetic CVs (dwarf nova and nova-like systems) 
and one class of the magnetic CVs, the IPs, are addressed. 
Both sub-classes have relatively well-understood X-ray spectra
which results in a relatively clean selection of objects.
The strongly magnetic CVs, the polars, will not be addressed
here. Polars are special due to their more complex X-ray spectra with a
thermal and a potentially strong soft component. ROSAT has uncovered many soft
polars~\citep{beuermann+schwope94}, but all new discoveries made with
XMM-Newton lack the pronounced soft component \citep[see e.g.][ and references therein]{webb+18}. 
It is therefore questionable if the observed
sample of polars may be regarded as representative of the parent population.

\begin{table*}
\caption{Non-magnetic CVs found in the RBS. \#713 is no longer considered non-magnetic, whereas 
\#664 was previously classified as an IP. The luminosity is given in the
  ROSAT spectral band 0.1--2.4\,keV. Distances and luminosities from
  ~\citet{schwope+02} are listed in columns with tilde, all other values are
  from this work. $V_{\rm gen}$ was computed for a scale height of 200\,pc.}
\label{t:nmcvs}
\begin{tabular}{rlccrrrrrrr}
RBS\# &Name        &  $\widetilde{D}$ & $\widetilde{\log L_X}$ & $f_{\rm X}$ &
$p$ & $\Delta p$ & $\log L_X$ & $D$ & $D_{\rm max}$ & $V_{\rm gen}$ \\
   &               & [pc]&  [s$^{-1}$] & [erg cm$^{-2}$ s$^{-1}$] & [mas] &
[mas] & [s$^{-1}$] & [pc]& [pc]& [pc$^{3}$] \\
\hline
   2 &EF Tuc       & 500:& 32.0&  3.38E-12&  0.720&  0.023&   32.9&  1335&  1462&   2.09E+08\\
  22 &WW Cet       & 100 & 31.0&  9.16E-12&  4.588&  0.047&   31.7&   216&   389&   3.28E+07\\
 280 &TT ARI       & 135 & 31.1&  5.96E-12&  3.884&  0.070&   31.7&   256&   372&   4.20E+07\\
 288 &WX HYI       & 265 & 31.6&  6.26E-12&  4.273&  0.029&   31.6&   232&   346&   3.21E+07\\
 372 &IQ Eri       & 130 & 30.8&  3.76E-12&  5.186&  0.167&   31.2&   192&   222&   1.11E+07\\
 490 &             &  33 & 29.6&  3.36E-12&  3.098&  0.126&   31.6&   320&   349&   3.51E+07\\
 512 &VW Hyi       &  65 & 30.9&  1.59E-11& 18.531&  0.022&   30.7&    54&   128&   3.26E+06\\
 664 &TW Pic       &  -- & --  &  3.13E-12&  2.284&  0.022&   31.8&   432&   455&   7.99E+07\\
 694 &SU UMa       & 280 & 32.1&  1.41E-11&  4.535&  0.029&   31.9&   219&   490&   9.25E+07\\
 710 &SW UMa       & 140 & 30.9&  3.81E-12&  6.148&  0.080&   31.1&   162&   188&   9.09E+06\\
(713 &EI UMa)      &  -- &   --&          &       &       &       &       &&\\
 728 &BZ UMa       & 110 & 30.9&  6.12E-12&  6.557&  0.064&   31.2&   152&   224&   1.39E+07\\
1008 &T Leo        & 100 & 31.1&  9.88E-12&  7.814&  0.069&   31.3&   128&   240&   1.33E+07\\
1411 &RHS40        & 460 & 32.0&  3.47E-12&  2.919&  0.671&   31.7&   367&   407&   3.72E+07\\
1900 &TY PsA       & 300:& 31.7&  4.68E-12&  5.431&  0.065&   31.3&   183&   236&   1.26E+07\\
1955 &V405 Peg     &  30 & 29.5&  3.05E-12&  5.784&  0.062&   31.0&   172&   179&   8.06E+06\\
1969 &CC Scl       & 165 & 31.1&  4.23E-12&  4.904&  0.148&   31.3&   203&   249&   1.37E+07\\
\hline
\end{tabular}
\end{table*}

\section{Analysis}
\subsection{Basic assumptions}
In the following we derive space densities for X-ray selected samples of
CVs. The samples are described in the literature, no new sample composition
was undertaken. Basic assumptions per case are:
\begin{itemize}
\item the observed sample is representative of the intrinsic population
\item the sample is complete
\item the sample is flux-limited with a well-defined flux-limit
\end{itemize}
For further details see e.g.~\citet{pretorius+knigge12}. 

For the analysis of the following sections we make use of parallaxes that were
found by archival cone searches in the Gaia archive using data from Gaia
DR2\footnote{\url{https://gea.esac.esa.int/archive/}}. In case of multiple matches
the entry with the best matching brightness value was chosen. We do not invert 
parallaxes to infer distances but use the probabilistic distance estimates 
provided by \citet{bailer-jones+18} that were accessed via \url{http://gaia.ari.uni-heidelberg.de/tap.html}. 
However, since most new parallaxes have very small relative errors the use of directly 
inverted parallaxes gives almost the same results.

\subsection{The $V/V_{\rm max}$ method}
We will follow the approach used earlier~\citep[e.g.][]{hertz+90,schwope+02,pretorius+07} to estimate the space
density of CVs using a $V/V_{\rm max}$ method. 
Since many of the CVs used here are at high galactic latitude and some of them
are at a distance in excess of the likely scale height of CVs, 
we use the modified method by~\citet{tinney+93}. This method 
of calculation of a generic volume, $V_{\rm gen}$, accounts for an exponential
density distribution  
$\rho \propto \exp[(-d |\sin b |)/h]$ 
($d$: distance, $b$: galactic latitude, $h$: galactic scale height).
$V_{\rm gen}$ is calculated by 
\begin{equation}
V_{\rm gen} = \Omega \frac{h^3}{|\sin^3b|}(2 - (\xi^2 + 2\xi +2)e^{-\xi})
\label{e:vgen}
\end{equation}
with $\xi = d |\sin b | / h$ and $\Omega$ the solid angle of the survey.
The maximum generic volume $V_{\rm gen}$ is computed using this formula with
the  maximum possible distance of the particular source which 
would allow its detection at the flux limit of the survey.
One particular CV then contributes
$1/V_{\rm gen}$ to the space density $\rho_X$, i.e.~the space density
is 
\begin{equation}
\rho_X = \sum\frac{1}{V_{\rm gen}}. 
\label{e:sum}
\end{equation}
The derived numbers per object are given in the tables below.

Per observed sample one needs to specify the flux limit, which determines the
maximum distance for an object still to be detected, the scale height of the
distribution, and the 
solid angle of the survey.

Using Gaussian distributed parallax and flux errors per object, 90\% confidence regions 
for the derived $\rho$ per sample and per assumed scale height were computed by running 
of order 50000 simulated mock samples per case. The results with their confidence 
ranges are listed in Tab.~\ref{t:dens}.

X-ray luminosities in this paper are always computed via 
$L_X = 4\pi D^2 F_X$ without any possible geometric correction factor. 
This leads to the revised luminosities given in the Tables below. 

\subsubsection{The RBS sample of non-magnetic CVs}
\label{s:rbs}

%which we assume to be $h = 200$\,pc, 
%solid angle($2\pi$ in our case)
The ROSAT-sample of non-magnetic CVs described in~\citet{schwope+02} was drawn
from the ROSAT Bright Survey~\citep[RBS,][]{schwope+00}, an identification
programme of all high-galactic latitude sources found in the RASS. It reached an identification rate as high as 99.7\%.
Based on the identification of two apparently
close, low-luminosity systems, RBS0490 and RBS1955,~\citet{schwope+02}
derived a space density of $\sim 3\times 10^{-5}$\,\ppc. Due to their
assumed proximity they were the dominating terms in the sum of
Eq.~\ref{e:sum}. When the two objects were removed from their sample, the
density became $\rho_{\rm X,RBS} = 1.5\times10^{-6}$\,\ppc. 

Triggered by this study,~\citet{thorstensen+06} and~\citet{thorstensen+09}
revised the distances to RBS0490 and RBS1955 to $\sim$300\,pc and
$149^{+26}_{-20}$\,pc, respectively, thus favouring the lower value of $\rho$ which was later confirmed by \citet{pretorius+knigge12}.
 
The original list of RBS-CVs is given in Table~\ref{t:nmcvs}. For a
re-determination of the space density some updates are necessary.
Firstly, RBS0713 (= EI UMa) is now regarded being an 
Intermediate Polar~\citep{baskill+05} and will not be included in the
analysis. On the other hand 
RBS0664 was regarded as an IP previously and was re-classified as a non-magnetic CV 
by \citet{pretorius+knigge12}.

Secondly, the RASS was recently reprocessed by~\citet{boller+16} and
count rates were updated. We read the revised count rates from the
online version of the
catalogue\footnote{\url{http://rosat.mpe.mpg.de/cgi-bin/2RXS-funcone-search}} 
and convert those to fluxes using the same ECF (energy to count conversion
factor) as in~\citet{schwope+02}, ECF = $1.41 \times 10^{-11}$\,\ergcmsqs / count.

Gaia distances are also listed in the Table. All but one 
have errors $<5$\%. There is one exception, RBS1411, 
with a relative uncertainty of 23\%.

The corresponding numbers for $V_{\rm gen}$ 
are listed in Table~\ref{t:nmcvs} together with the X-ray flux, the parallax 
(plus error), the estimated distance and the maximum distance 
that was used to compute the generic volume. The survey area used was 20400\,deg$^2$. 

Here and in the other subsections the generic volume
was computed for three different scale heights, those that were used in the 
original publications, 200\,pc for the RBS-CVs, 260\,pc for the RASS-CVs and 120\,pc 
for the IPs. 
All derived space densities are listed in Tab.~\ref{t:dens}.

\begin{table*}
\caption{Non-magnetic CVs used by~\citet{pretorius+knigge12} to obtain the
  space density of RASS-selected objects. The luminosity 
is given in the spectral band 0.5--2.0\,keV. Distances
and luminosities with tilde are from the original publication, those without 
from this work. $V_{\rm gen}$ was computed for a scale height of 260\,pc.}
\label{t:pret}
\begin{tabular}{rlccrrrrrrr}
RBS\# &Name        &  $\widetilde{D}$ & $\widetilde{\log L_X}$ & $f_{\rm X}$ &
$p$ & $\Delta p$ & $\log L_X$ & $D$ & $D_{\rm max}$ & $V_{\rm gen}$ \\
   &               & [pc]&  [s$^{-1}$] & [erg cm$^{-2}$ s$^{-1}$] & [mas] &
[mas] & [s$^{-1}$] & [pc]& [pc]& [pc$^{3}$] \\
\hline
   2 & EF Tuc    &  346 & 31.4 & 1.60E-12&  0.720&  0.023&   32.5&  1335&  1610&   4.25E+08\\ 
  22 & WW Cet    &  158 & 31.2 & 5.70E-12&  4.588&  0.047&   31.5&   216&   492&   6.84E+07\\
 280 & TT Ari    &  335 & 31.7 & 3.50E-12&  3.884&  0.070&   31.4&   256&   457&   8.16E+07\\
 288 & WX Hyi    &  260 & 31.4 & 3.00E-12&  4.273&  0.029&   31.3&   232&   383&   5.03E+07\\
 372 & IQ Eri    &  116 & 30.4 & 1.20E-12&  5.186&  0.167&   30.7&   192&   201&   1.01E+07\\
 490 &           &  285 & 31.2 & 1.60E-12&  3.098&  0.126&   31.3&   320&   386&   5.42E+07\\
 512 & VW Hyi    &   64 & 30.5 & 6.10E-12& 18.531&  0.022&   30.3&    54&   127&   3.40E+06\\
 664 & TW Pic    &  230 & 31.0 & 1.70E-12&  2.284&  0.022&   31.6&   432&   537&   1.42E+08\\
 694 & SU UMA    &  261 & 31.9 & 9.00E-12&  4.535&  0.029&   31.7&   219&   626&   1.97E+08\\
 710 & SW UMa    &  164 & 30.6 & 1.40E-12&  6.148&  0.080&   30.6&   162&   183&   9.24E+06\\
 728 & BZ UMa    &  228 & 31.1 & 2.30E-12&  6.557&  0.064&   30.8&   152&   220&   1.49E+07\\
1008 & T Leo     &  101 & 30.6 & 3.60E-12&  7.814&  0.069&   30.8&   128&   232&   1.45E+07\\
1411 & RHS40     &  468 & 31.6 & 1.90E-12&  2.919&  0.671&   31.5&   367&   482&   6.92E+07\\
1900 & TY PsA    &  239 & 31.3 & 2.80E-12&  5.431&  0.065&   31.1&   183&   292&   2.48E+07\\
1955 & V405 Peg  &  149 & 30.5 & 1.20E-12&  5.784&  0.062&   30.6&   172&   179&   8.92E+06\\
1969 & CC Scl    &  359 & 31.4 & 1.50E-12&  4.904&  0.148&   30.9&   203&   237&   1.48E+07\\
NEP1 & EX Dra    &  240 & 29.7 & 8.10E-14&  4.058&  0.019&   29.8&   244&   491&   4.93E+05\\
NEP2 & RXJ1831   &  980 & 31.5 & 2.80E-13&  1.232&  0.038&   31.3&   793&  2967&   8.60E+06\\
NEP3 & SDSSJ1730 &  444 & 31.2 & 7.30E-13&  1.864&  0.038&   31.4&   528&  3190&   5.08E+06\\
NEP4 & RXJ1715   &  400 & 30.4 & 1.20E-13&  1.382&  0.146&   30.9&   713&  1747&   3.64E+06\\
\hline
\end{tabular}
\end{table*}

\begin{table*}
\caption{Intermediate Polars used by~\citet{pretorius+mukai14} to obtain the
  space density. The luminosity is given in the spectral band 14 --
  195\,keV. Distances and luminosities with tilde are from the orginal
  publication, those without are from this work. $V_{\rm gen}$ was 
computed for a scale height of 120\,pc.}
\label{t:ips}
\begin{tabular}{lccrrrrrrr}
Name        &  $\widetilde{D}$ & $\widetilde{\log L_X}$ & $f_{\rm X}$ &
$p$ & $\Delta p$ & $\log L_X$ & $D$ & $D_{\rm max}$ & $V_{\rm gen}$ \\
            & [pc]&  [s$^{-1}$] & [erg cm$^{-2}$ s$^{-1}$] & [mas] &
[mas] & [s$^{-1}$] & [pc]& [pc]& [pc$^{3}$] \\

%Name        &  $\widetilde{D}$ & $\widetilde{\log L_X}$ &
%$D$ & $\log L_X$ & $V_{\rm act}$ & $V_{\rm gen}$ & $V_{\rm act}/V_{\rm gen}$
%& $ 1 / V_{\rm gen}$\\
%            & [pc]&  [s$^{-1}$] & [pc]&  [s$^{-1}$] & [pc$^{3}$] &
%[pc$^{3}$] & & [pc$^{-3}$]\\
\hline
V1223 Sgr  &    527 & 33.6 & 1.18E-10&  1.725&  0.047&   33.7&   571&  1241&   1.22E+09\\
V2400 Oph  &    280 & 32.7 & 5.00E-11&  1.399&  0.033&   33.5&   701&   991&   1.51E+09\\
AO Psc     &    330 & 32.6 & 3.20E-11&  2.021&  0.043&   33.0&   488&   552&   5.48E+07\\
IGR J16500 &    430 & 32.8 & 2.60E-11&  0.855&  0.062&   33.6&  1140&  1163&   2.46E+09\\
V405 Aur   &    380 & 32.8 & 3.50E-11&  1.487&  0.031&   33.3&   662&   783&   5.78E+08\\
FO Aqr     &    450 & 33.1 & 5.20E-11&  1.902&  0.051&   33.2&   518&   747&   7.76E+07\\
PQ Gem     &    510 & 33.0 & 3.10E-11&  1.306&  0.036&   33.3&   750&   835&   4.28E+08\\
TV Col     &    368 & 33.0 & 6.00E-11&  1.951&  0.018&   33.3&   505&   782&   1.94E+08\\
IGR J15094 &    960 & 33.5 & 2.60E-11&  0.859&  0.028&   33.6&  1127&  1149&   2.35E+09\\
XY Ari     &    270 & 32.5 & 3.60E-11&  3.700&  1.500&   32.5&   270&   324&   4.33E+07\\
EI UMa     &    750 & 33.3 & 2.90E-11&  0.883&  0.037&   33.6&  1095&  1179&   1.66E+08\\
NY Lup     &    680 & 33.7 & 9.20E-11&  0.786&  0.028&   34.2&  1228&  2356&   9.38E+09\\
V1062 Tau  &   1400 & 33.8 & 2.50E-11&  0.632&  0.077&   33.8&  1512&  1512&   2.72E+09\\
V2731 Oph  &   1000 & 33.9 & 6.90E-11&  0.436&  0.056&   34.6&  2165&  3597&   2.25E+09\\
GK Per     &    477 & 33.3 & 7.80E-11&  2.263&  0.043&   33.3&   437&   772&   7.74E+08\\
\hline
\end{tabular}
\end{table*}

A few of the non-magnetic RBS-CVs might have an uncertain classification hence
a final composition of the sample is subject to changes if newer information
becomes available. 
One example is RBS1955 which is difficult to classify, it could well be a magnetic CV~\citep{schwope+14}. 
If removed from the sample one obtains a density 10\% lower than that given in Tab.~\ref{t:dens}.

\subsubsection{The RASS (RBS \& NEP) sample of non-magnetic CVs}\label{s:rass}
\citet{pretorius+knigge12} used the non-magnetic RBS-CVs and added four CVs
from the ROSAT-NEP \citep[North Ecliptic Pole, ][]{pretorius+07} 
survey to study the space density and 
the X-ray luminosity  function of CVs.
They re-considered all distance determinations
used previously to obtain  $\rho_X = 4^{+6}_{-2}\times10^{-6}$\,\ppc\ for an
assumed scale height of 260\,pc. Their error budget is based on Monte-Carlo
simulations of the probability distribution function that find $\rho$ for a
large number of mock samples with properties that fairly sample the parameter
space allowed by the data.

The list of objects with their newly determined distances, luminosities and other parameters
is given in Table~\ref{t:pret}. The survey area used is 20400\,deg$^2$ at a limiting flux of 
$1.1\times 10^{-12}$\,erg cm$^{-2}$ s$^{-1}$
for the RBS part and 81\,deg$^2$ for the NEP part of the sample. 

As a test for consistency we re-calculated the space density using their data
as far as we were able to recover those. The limiting flux is not constant
over the NEP area and \cite{pretorius+07} describe how to correctly deal with
the variable flux limit. It  
is not expected that the correct treatment makes a significant difference to
the results achieved here. 
Following~\citet{henry+06} we thus simply used the same limit of
$2\times10^{-14}$\,\ergcmsqs\ for the NEP survey area to
obtain $\rho_{\rm X, RASS} = 4.1\times10^{-6}$\,\ppc\, as a reference value assuming the same 
scale height of $h=260$\,pc\ as in~\citet{pretorius+knigge12}.

\cite{pretorius+knigge12} used a slightly different X-ray band as
\citet{schwope+02} and corrected their fluxes for interstellar extinction
which explains different derived values of $\rho_X$ despite using the same
objects. As we will show below, the differences are small compared to other
parameters affecting $\rho$. 
We thus  tested the influence of the sample composition and different flux
convention used by \cite{schwope+02} and \cite{pretorius+knigge12} by removing
the 4 NEP-CVs from the RASS-sample and re-computing the space density. The
results are listed in Tab.~\ref{t:dens} in the row labeled 'RASS (RBS-part)'.

\subsubsection{The sample of intermediate polars from the Swift/BAT 70 month
  survey}\label{s:ips}
The third sample to be studied here is the Swift/BAT sample of IPs presented
by \citet{pretorius+mukai14}. They list 15 IPs that were observed in the energy
range 14--195\,keV. This band is not affected by galactic foreground absorption. 
The limiting flux of this survey is $F_X = 2.5\times
10^{-11}$\,\ergcmsqs\, the survey was restricted to galactic latitudes 
$b^{\rm II} > |5|\degr$. The space density derived for long-period IPs with an assumed
scale height of 120\,pc was $\rho_X = 1^{+1}_{-0.5} \times10^{-7}$\,\ppc. 

The IPs used in this exercise  with their newly determined distances and
luminosities are listed in Table~\ref{t:ips}.

XY Ari is behind a dark cloud, it has no optical counterpart and is therefore
without data from Gaia. We use the same distance as~\citet{pretorius+mukai14}. 
V2731 Oph has a relative parallax error of 13\%, and V1062 Tau a relative
error of 12\%. Most other parallax errors are below 3\%.  

\begin{table*}
\caption{Space densities of the X-ray selected CV samples as a function of assumed scale height. Errors are given at 90\% confidence
\label{t:dens}
}
\begin{tabular}{lrrr}
& 120\,pc & 200\,pc & 260\,pc\\
\hline\\[-1ex]
RBS   & 
$1.62_{-0.06}^{+0.14} \times 10^{-6}$ & 
$1.09_{-0.05}^{+0.11} \times 10^{-6}$ & 
$0.96_{-0.05}^{+0.10} \times 10^{-6}$ 
\\[1.5ex]
RASS  & 
$1.09_{-0.05}^{+0.09} \times 10^{-5}$ & 
$4.71_{-0.37}^{+0.66} \times 10^{-6}$ & 
$3.56_{-0.33}^{+0.58} \times 10^{-6}$ 
\\[1.5ex]
RASS (RBS-part) & 
$1.60_{-0.08}^{+0.17} \times 10^{-6}$ & 
$1.08_{-0.06}^{+0.14} \times 10^{-6}$ & 
$0.95_{-0.06}^{+0.13} \times 10^{-6}$ 
\\[1.5ex]
Swift &
$7.4_{-1.7}^{+4.8} \times 10^{-8}$ & 
$3.6_{-1.3}^{+4.0} \times 10^{-8}$ & 
$2.8_{-1.2}^{+3.7} \times 10^{-8}$ 
\\[0.5ex]
\hline
\end{tabular}
\end{table*}

\begin{figure}
\resizebox{\hsize}{!}{\includegraphics[clip=]{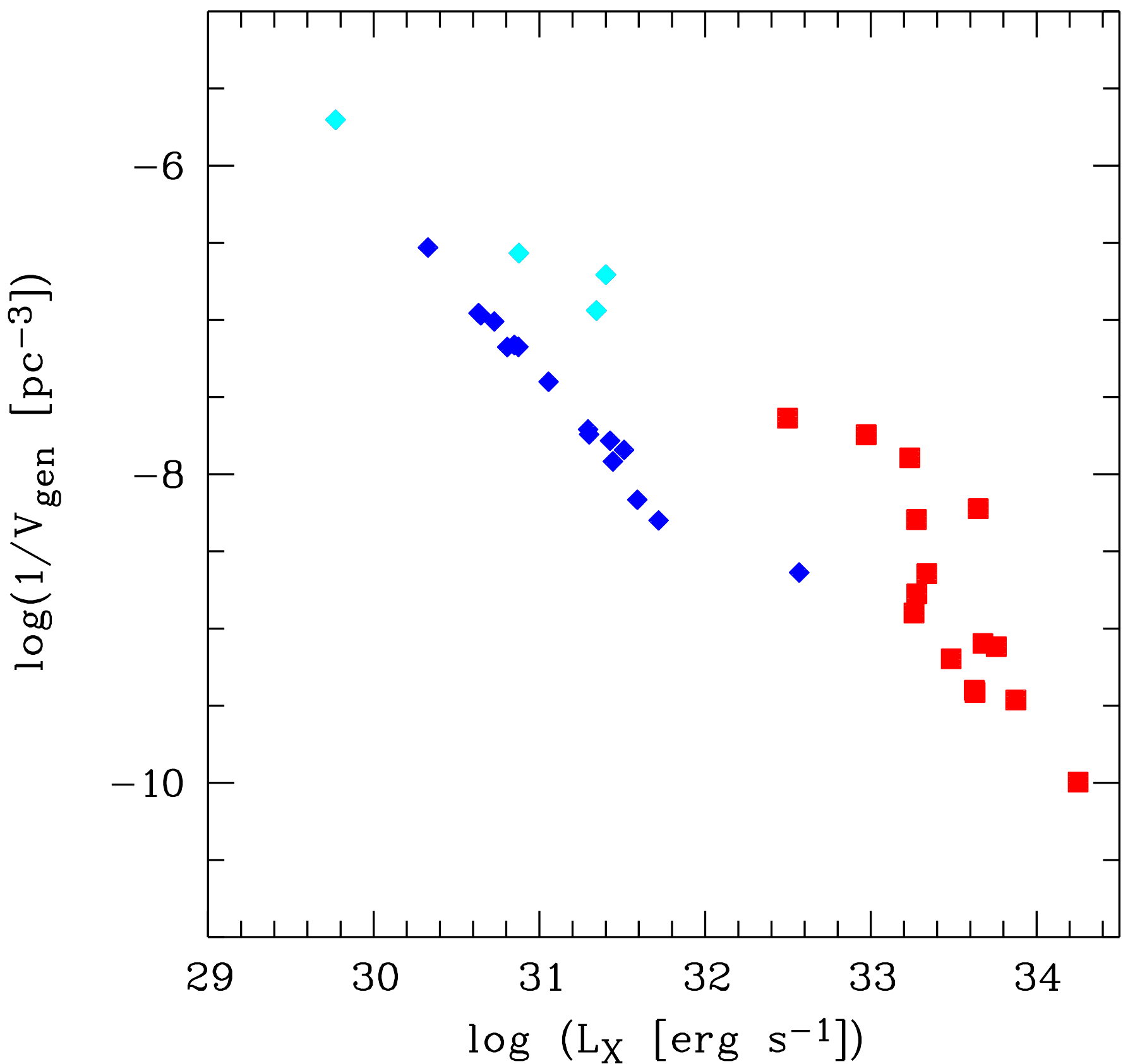}}
\caption{Scatter plot showing the X-ray luminosities and the 
weights (inverse of the generic volumes) per object studied here. 
The Swift/BAT selected IPS are shown in red for a scale height 
of 120\,pc, the RASS CVs in blue, the NEP-CVs in the RASS sample 
in cyan. Assumed scale height for RASS-CVs was 260\,pc.
\label{f:lx_wei}}
\end{figure}

\begin{figure}
\resizebox{\hsize}{!}{\includegraphics[clip=]{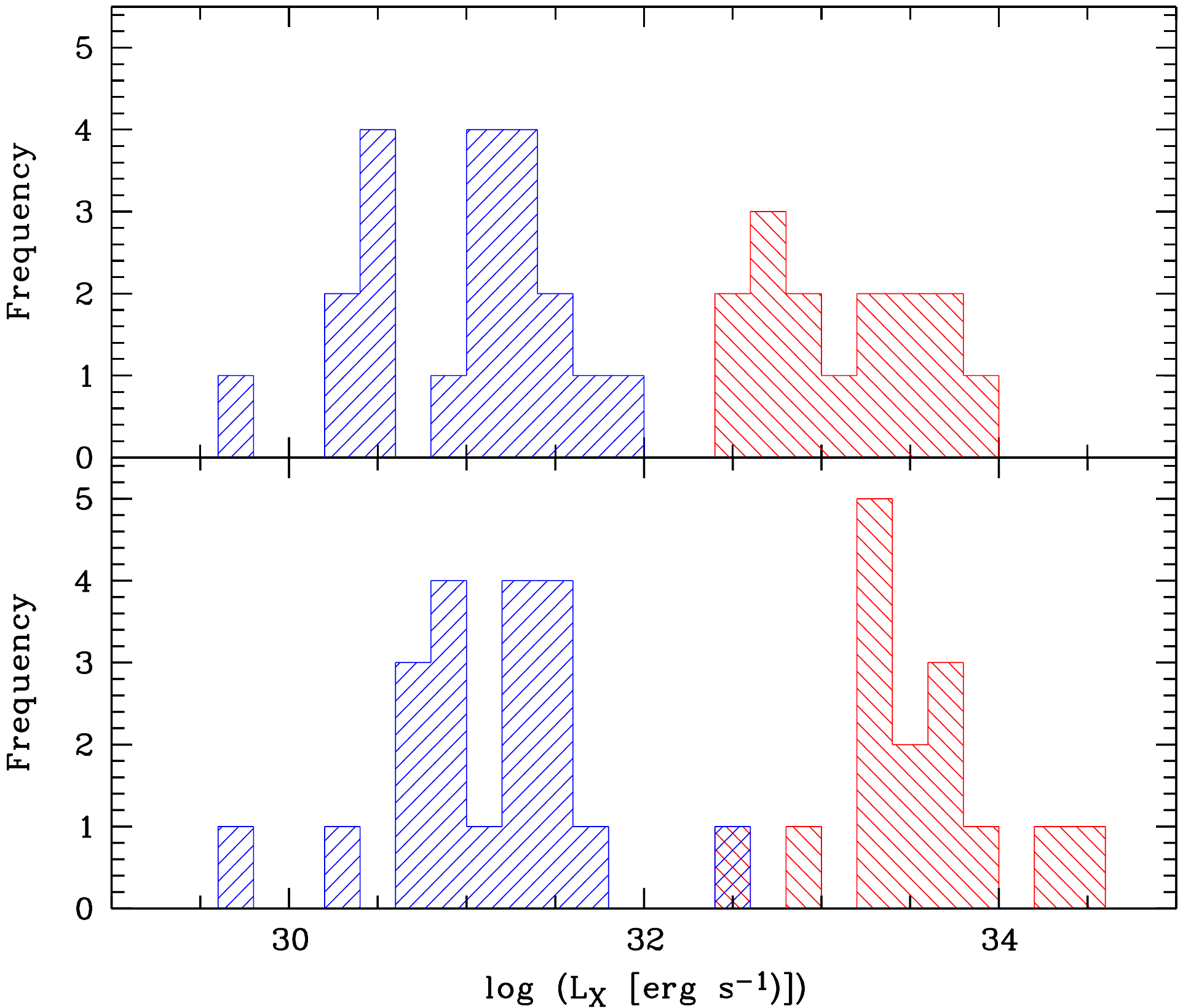}}
\caption{Published \citep[upper panel, adapted from ][]{pretorius+knigge12,pretorius+mukai14} and revised (lower panel)
luminosity distributions of the RASS-CVs (blue-shaded histogram) 
and the Swift/BAT-selected IPs (red). The bin width is 0.2 dex.
\label{f:lxdist}}
\end{figure}

\section{Results  and discussion}
\label{s:results+discussion}
The space densities of X-ray selected samples of magnetic and non-magnetic CVs 
was redetermined using recently published parallaxes and distances from Gaia-DR2. 
The results are summarized in Tab.~\ref{t:dens} and Figs.~\ref{f:lx_wei} and \ref{f:lxdist}.
Fig.~\ref{f:lxdist} shows distributions of the original and the revised luminosities
found for the RASS-CVs and the IPs, while Fig.~\ref{f:lx_wei} shows the weight per object 
(inverse of the generic volume) over its luminosity.

Just as a reference, the published values of $\rho$ for the RBS, 
the RASS and the IP samples were $\sim1.5\times10^{-6}$\,\ppc, 
$4_{-2}^{+6} \times 10^{-6}$\,\ppc, 
and $1_{-0.5}^{+1} \times 10^{-6}$\,\ppc, 
for scale heights of 200\,pc, 260\,pc, and 120\,pc, respectively {\bf \citep{schwope+02,pretorius+knigge12,pretorius+mukai14}}. 
The comparison with the values listed in Tab.~\ref{t:dens} shows
that all newly derived densities are smaller than published ones 
but still within the published errors.

The statistical errors of $\rho$ of the non-magnetic CVs 
could be reduced very significantly thanks to precise Gaia data.
The statistical error of $\rho$ for the IP sample is still large
due to the shallow flux limit.

For the RBS-CVs the space density is safely below $1.8 \times 10^{-6}$\,\ppc\
at 90\% confidence but could be smaller than $1.1 \times 10^{-6}$\,\ppc\ if
the scale height would be as high as 260\,pc as assumed by \cite{pretorius+knigge12}.
This result remains unchanged if the fluxes in the band $0.5-2.0$\,keV
are used (third row in Tab.~\ref{t:dens}). The RBS-sample consists 
of long- and short period CVs. It thus appears possible that not all objects belong 
to the same galactic population. The use of just one scale to characterise 
the sample is likely an oversimplification. 

The RASS-CVs (RBS$+$NEP) are compatible with a significantly higher space
density thanks to the lower flux limit of the NEP. The inclusion of just 4 CVs from the NEP-survey implies a space 
density a factor 4 to 7 larger than without those. Fig.~\ref{f:lx_wei}
illustrates that at a given luminosity each of those CVs has a factor 
$\sim$10 higher weight than a corresponding RBS-CV. The whole sample is 
dominated by just one CV, the low-luminosity object EX Dra, 
$\log L_X = 29.8$\,\ergs, a rather unhealthy situation for the whole 
analysis. 

The pre-Gaia distances of the RASS-CVs were quite reliable, hence 
the median X-ray luminosity of the RASS sample 
remained unchanged at $\log L_X = 31.2$\,\ergs. The distribution 
of luminosities is less dispersed in the center but a bit more fuzzy 
at the outskirts (Fig.~\ref{t:dens}). 
The standard deviation of $\log L_X$ was 0.54\,dex
and is now 0.57\,dex, omitting the highest and lowest values it was 0.46\,dex
and is now 0.38\,dex.

For the Intermediate Polars the first important thing to note is that all but 
one object, GK Per, have larger distances than previously assumed. 
Hence, they are more luminous then thought, the median luminosity 
is shifted from $\log L_X = 33.1$\,\ergs\ to $33.5$\,\ergs, the survey volume 
becomes larger and the space density conversely smaller. An upper limit to the space density 
of the IPs is $\rho < 1.3 \times 10^{-7}$\,\ppc\ at 90\% confidence, a 
significant reduction compared to the published value. The most likely 
value at $7.4_{-1.7}^{+4.8} \times 10^{-8}$\ppc\ is at 74\% of the published one.

Distances to the IPs used by \cite{pretorius+mukai14} 
were either taken from the literature (3 trigonometric
and 4 photometric parallaxes from the donor star) or were 
newly determined and based on WISE IR-magnitudes combined 
with the semi-empirical donor sequence by \cite{knigge06}.
Not surprisingly, the mean distance ratio new/old is reasonably 
small, $d_{\rm rat} = 1.12$ for the 
three IPs which had a trigonometric parallax previously, among them 
GK Per with a Gaia distance smaller than published. The IPs with 
photometric parallaxes of the donor have $d_{\rm rat} = 1.33$ and 
those with estimated distances from WISE and the empirical donor sequence
have a mean ratio $d_{\rm rat} = 1.74$.
This leaves the two possibilities that either the IR donor sequence is 
somehow biased or that an additional emission component (dust, cyclotron 
radiation, free-free emission) mimics brighter secondaries.

Otherwise the IP sample appears more homogeneous than the sample of non-magnetic CVs.
There is not one object or a subgroup of objects that dominates the space density. 
However, given the rather high flux limit the number for $\rho$ derived here 
is valid only for the potentially rare objects with high luminosities. 
The putative class of low-luminosity IPs remains yet to be uncovered 
\citep[see e.g.~][]{worpel+18}.

In this study an update on the space density of X-ray selected CVs was given.
CVs appear to be rarer than previously thought. While the limitations due to 
uncertain distances are overcome thanks to Gaia major obstacles preventing 
further progress remain. These are the small sample sizes due to shallow 
flux limits of past X-ray surveys and the ignorance about the proper scale 
heights of the samples. It also appears likely that the existing samples are
inhomogeneously composed as far as their scale height is concerned, 
they contain long-and short-period objects. 
Part of them lack determinations of their orbital period, which could be 
used to assign class membership, belonging to an older or younger population 
with corresponding scale height.

The limitation given by the small sample sizes will hopefully soon be 
overcome as a result of the upcoming eROSITA all-sky surveys~\citep{merloni+12,schwope12}
with an all-sky flux limit comparable to the ROSAT-NEP survey but with enlarged 
energy coverage, $0.3-10$\,keV, and better spatial resolution compared to ROSAT.
Performing the survey is just the first step on a longer ladder which will 
involve spectroscopic identification, classification and detailed follow-up 
to determine orbital periods. 

\begin{acknowledgements}
I thank Fabian Emmerich for help with type-setting and Hauke W\"orpel for 
useful hints regarding Gaia distances.

I thank an anonymous referee whose comments helped to improve the quality and clarity of the paper.

This work has made use of data from the European Space Agency (ESA)
 {\it Gaia} (\url{https://www.cosmos.esa.int/gaia}), processed by
the {\it Gaia} Data Processing and Analysis Consortium (DPAC,
\url{https://www.cosmos.esa.int/web/gaia/dpac/consortium}). Funding
for the DPAC has been provided by national institutions, in particular
the institutions participating in the {\it Gaia} Multilateral Agreement.

\end{acknowledgements}

\bibliographystyle{aa}
\bibliography{rhox}

\end{document}